\documentclass[aps,amssymb,floatfix,prd,amsmath,preprintnumbers]{revtex4}
\usepackage{epstopdf}
\usepackage{capt-of}
\usepackage{float}
\usepackage{graphicx} 
\usepackage{dcolumn} 
\usepackage{hyperref} 
\usepackage{bm}
\begin{document}
\title{Dark Energy and Modified Scale Covariant Theory of Gravitation}

\author{Koijam Manihar Singh\footnote{School of Engineering and Technology, Mizoram University, Aizawl-796004, India, Email: drmanihar@rediffmail.com},
Sanjay Mandal\footnote{Department of Mathematics, Birla Institute of Technology and Science-Pilani, Hyderabad Campus, Hyderabad-500078, India, Email: sanjaymandal960@gmail.com},
Longjam Parbati Devi \footnote{N. G. College, Lamphelpat, Imphal-795004, India, Email: parbatilongjam22@gmail.com},
      P.K. Sahoo\footnote{Department of Mathematics, Birla Institute of Technology and Science-Pilani, Hyderabad Campus, Hyderabad-500078, India, Email: pksahoo@hyderabad.bits-pilani.ac.in}}

\affiliation{ }

\begin{abstract}
\textbf{Abstract:} Taking up four model universes we study the behaviour and contribution of dark energy to the accelerated expansion of the universe, in the modified scale covariant theory of gravitation. Here, it is seen that though this modified theory may be a cause of the accelerated expansion it cannot totally outcast the contribution of dark energy in causing the accelerated expansion. In one case the dark energy is found to be the sole cause of the accelerated expansion. The dark energy contained in these models come out to be of the $\Lambda$CDM type and quintessence type comparable to the modern observations. Some of the models originated with a big bang, the dark energy being prevalent inside the universe before the evolution of this era. One of the models predicts big rip singularity, though at a very distant future. It is interestingly found that the interaction between the dark energy and the other part of the universe containing different matters is enticed and enhanced by the gauge function $\phi(t)$ here.

\end{abstract}

\maketitle

\section{Introduction}
The observational data such as type Ia supernovae, cosmic microwave background (CMB), and baryonic acoustic oscillations (BAO) confirm that the universe is undergoing accelerated expansion \cite{Riess/1998, Perlmutter/1999, Tegmark/2004, Eisenstein/2005, Spergel/2003}, which brings revolution to the modern cosmology and answer many questions related to the evolution of the universe. However we know very little about the accelerated expansion of the universe. Approximately sixty-eight percent of the universe \cite{Ade/2016} is covered by a mysterious component so-called ``dark energy" which is associated with negative pressure. Many researchers believe that dark energy is responsible for the accelerated expansion of the universe. Dark energy is one suitable gradient for the accelerated expansion of the universe, because it has a tendency to overcome the collapsing nature of the matter in the universe due to gravity, and also provides some informations about the accelerated expansion of the universe \cite{Weinberg/ 2008}.\\
There are some commonly accepted ways for describing the accelerated expansion of the universe. These are through modified theories of gravity and also considering different kinds of cosmological fluids \cite{Nojiri/2011, Bamba/2012, Dvali/2000, Yin/2007}, holographic dark energy \cite{Wang/2006, Nojiri/2006}, cosmological constant, some scalar or vector fields \cite{Padmanabhan/2002, Bagla/2003, Brax/1999, Copeland/2000}, quintessence \cite{Peebles/1988, Ratra/1988, Frieman/1985, Caldwell/1998}, meta-stable dark energy \cite{Stojkovic/2008, Greenwood/2009}, etc.\\
To describe the accelerated expansion of the universe using the theory of general relativity, we need dark energy in the universe which is valid on all scales of the universe. After the discovery of cosmic acceleration, the modified gravitation theories play an important role in describing the accelerated expansion of the universe. Mention may be made of $f(R)$ gravity, scalar tensor theories, Galilean gravity, Gauss-Bonnet gravity, braneworld models and so on \cite{Carroll/2004, Nojiri/2003, Amendola/1999, Uzan/1999, Chiba/1999, Bartolo/2000, Perrotta/2000, Riazuelo/2002, Dvali/2000, Nicolis/2009, Nojiri/2005}. Higher dimensional Kaluza-Klein space-time with perfect fluid source was studied in \cite{Joey/2016, Luca/1990}, and the dark energy model with variable deceleration parameter with the same metric was investigated in the scale covariant theory of gravitation \cite{Reddy/2018}. The spatially homogeneous Bianchi type dark energy models were discussed with different equations of state by some authors in this theory \cite{Anirudh/2011, Adhav/2012}. Kantowski-Sachs anisotropic dark energy models and bulk viscous string cosmological models were considered in \cite{Pawar/2014, Katore/2012}.\\
The $\Lambda$CDM model is the standard cosmological model which helps to explain of the universe, where $\Lambda$ stands for vacuum energy and CDM stands for cold dark matter are mysterious. The dark sectors are occupied $95$\% of the total energy of the universe. There is possibility of momentum exchange or energy exchange or the interaction between the dark sector components. So, researchers  can't ignore these when they are trying to study the current phenomenon of the universe. As a result, a large number of studies have been proposed to carry out the interaction between the dark sectors of the universe with different prospectives and motivations \cite{41, 42, 43, 44, 45, 46, 47, 48, 49, 50, 51, 52, 53, 54}. The $\Lambda$CDM model extended by allowing the interaction between the dark sector components of the universe with the observational datas like Plank-CMB, KiDs, HST, BAO \cite{42, 43, 44, 45, 54}.\\
 In our problem we have considered modification of Einstein's gravitation theory as proposed by Canuto et al. (1977) \cite{Canuto/1977} , and also we have introduced some new interaction term $Q$ in its field equations. Our motivation is to study the accelerated expansion of the universe in its fundamental level by making changes in the scalar field with introduction of some new interaction terms.\\
Our problem is organised as follows. In section II, we have considered the modified gravitation theory proposed by Canuto et al. and construct the field equations by considering spherically symmetric metric, as investigation of the behaviour of dark energy in a spherically symmetric universe is interesting \cite{Koijam/2009}, and we draw the graphs showing the behaviour of energy density and pressure in each case. Section III is the discussion and conclusion of the different results obtained in these cases. 
\section{Formulation of the problem}
A modification of the Einstein's gravitation theory was proposed by Canuto et al. (1977) \cite{Canuto/1977} in which the gravitational constant is a variable and where gravitational units are used for Einstein's field equations, but atomic units are used for physical quantities. A conformal transformation bridges the two systems of units in the form
\begin{equation}
\label{eqn:1}
g_{ij}'=\phi^2(x^l)g_{ij}
\end{equation}
where $i, j, l=1, 2, 3,$ and dashed ones are gravitational units and undashed ones denote atomic units. Here it may be noted that the gauge function $\phi(0<\phi<\infty)$ is a function of space and time coordinates in the most general form. With the above transformation the modified form of the field equations due to Canuto et al. (1977) \cite{Canuto/1977} is
\begin{equation}
\label{eqn:2}
R_{ij}-\frac{1}{2}g_{ij}R+f_{ij}(\phi)=-8\pi GT_{ij}+\Lambda (\phi)g_{ij}
\end{equation}
with 
\begin{equation}
\label{eqn:3}
\phi^2f_{ij}=2\phi\phi_{i;j}-4\phi_{,i}\phi_{,j}-g_{ij}(\phi\phi_{;k}^{,k}-\phi^{,k}\phi_{,k})
\end{equation}
where the scalar $\phi$ is not accompanied with a wave equation. Thus Canuto et al. (1977) \cite{Canuto/1977} considered the most possible expressions for $\phi$ as
\begin{equation}
\label{eqn:4}
\phi=\gamma t^{-\alpha},  \alpha=\pm\frac{1}{2}, \pm 1
\end{equation}
where `$\gamma$' is a constant.
However Canuto and Goldman (1983) \cite{Canuto/1983} considered the relation
\begin{equation}
\label{eqn:5}
\phi\sim t^{\frac{1}{2}}
\end{equation}
as the most appropriate one in agreement with observations.
We consider here the FRW metric
\begin{equation}
\label{eqn:6}
ds^2=dt^2-a^2(t)(dx^2+dy^2+dz^2)
\end{equation}
where $a(t)$ is the scale factor of the universe.\\
The energy momentum tensor of the fluid contained in this universe is taken as
\begin{equation}
\label{eqn:7}
T_{ij}=(\rho_d+\rho_m+p_d+p_m)u_iu_j-(p_d+p_m)g_{ij}
\end{equation}
where $\rho_d$ and $p_d$ are respectively the energy density and fluid pressure for the dark energy contained, and $\rho_m$ and $p_m$ are respectively the energy density and pressure of other matters including dark matter.\\
Here $u^i=(1,0,0,0)$ is the four velocity flow vector satisfying the conditions
\begin{equation}
\label{eqn:8}
u^iu_i=1, u^iu_j=0
\end{equation}
Considering the co-moving coordinate system and taking $\Lambda=0$, the scale co-variant field equations take the forms
\begin{equation}
\label{eqn:9}
3H^2+3H\frac{\dot{\phi}}{\phi}-\frac{\ddot{\phi}}{\phi}+3\biggl(\frac{\dot{\phi}}{\phi}\biggr)^2=8\pi G(\phi)(\rho_d+\rho_m)
\end{equation}
\begin{equation}
\label{eqn:10}
2\dot{H}+3H^2+H\frac{\dot{\phi}}{\phi}+\frac{\ddot{\phi}}{\phi}-\biggl(\frac{\dot{\phi}}{\phi}\biggr)^2=-8\pi G(\phi)(p_d+p_m)
\end{equation}
\begin{equation}
\label{eqn:11}
3\dot{H}+6H^2+3H\frac{\dot{\phi}}{\phi}+\frac{\ddot{\phi}}{\phi}-\biggl(\frac{\dot{\phi}}{\phi}\biggr)^2=-8\pi G(\phi)(p_d+p_m),
\end{equation}
where $H=\frac{\dot{a}}{a}$.
Here the conservation equation takes the form
\begin{equation}
\label{eqn:12}
(\dot{\rho_d}+\dot{\rho_m})+(\rho_d+\rho_m+p_d+p_m)3H\\
+(\rho_d+\rho_m)\biggl(\frac{\dot{G}}{G}+\frac{\dot{\phi}}{\phi}\biggr)+3(p_d+p_m)\biggl(\frac{\dot{\phi}}{\phi}\biggr)=0
\end{equation}
which can be separated into two equations, namely;
\begin{equation}
\label{eqn:13}
\dot{\rho_d}+(\rho_d+p_d)3H+\rho_d\biggl(\frac{\dot{G}}{G}+\frac{\dot{\phi}}{\phi}\biggr)+3p_d\biggl(\frac{\dot{\phi}}{\phi}\biggr)=Q
\end{equation}
and\begin{equation}
\label{eqn:14}
\dot{\rho_m}+(\rho_m+p_m)3H+\rho_m\biggl(\frac{\dot{G}}{G}+\frac{\dot{\phi}}{\phi}\biggr)+3p_m\biggl(\frac{\dot{\phi}}{\phi}\biggr)=-Q
\end{equation}
{where $Q$ is the interaction term which describes the energy flow rate between dark matter and dark energy/vacuum energy. Here, $Q<0$ represents the decaying of energy from dark matter to dark energy, $Q=0$ represents no interaction between dark sectors, and $Q>0$ represents the decaying of energy from from dark energy to dark matter \cite{54, 55, 56, 57}. Generally, in literature the interaction is assumed as directly proportional to the density of dark sectors i.e. $Q \propto \rho $, where$\rho$ be the density of dark energy or dark matter or the combination of both dark energy and dark matter \cite{58, 59, 60, 61}. In this article, we use $Q=\beta H \rho_m$ and $Q=\beta H \rho_d$, where $\rho_m$ is the density of dark matter and $\rho_d$ is the density of dark energy.\\
Now from Eqns. (\ref{eqn:10}) and (\ref{eqn:11}) we have
\begin{equation}
\label{eqn:15}
\dot{H}+3H^2+2H\frac{\dot\phi}{\phi}=0
\end{equation}
In the following the equation of state parameter between the energy density $\rho_d$ and the pressure $p_d$ of dark energy is taken to be $\omega$ so that
\begin{equation}
\label{eqn:16}
p_d=\omega\rho_d
\end{equation}\\
\textbf{Case I :}\\ \\
The interaction term can take different forms, and here we take
\begin{equation}
\label{eqn:17}
Q=\beta H\rho_d
\end{equation}
where $H$ is the mean Hubble parameter and $\beta$ is a constant.\\
Now we consider the cases $\alpha=\frac{1}{2},\pm 1$ and $\alpha\sim\frac{-1}{2}$ in (\ref{eqn:4}); then equation (\ref{eqn:15}) gives
\begin{equation}
\label{eqn:18}
a(t)=\left\lbrace 3a_0\gamma^{-2}\left(\frac{t^{2\alpha+1}}{2\alpha+1}\right)+a_1\right\rbrace^{\frac{1}{3}}
\end{equation}
where $a_0$ and $a_1$ are arbitrary constants.\\
In this case we get, from equations (\ref{eqn:4}, \ref{eqn:17}, \ref{eqn:18}, \ref{eqn:13}),
\begin{equation}
\label{eqn:19}
\rho_d=\frac{a_2}{G}(\gamma t^{-\alpha})^{-3\omega-1}\left\lbrace 3a_0\gamma^{-2}\left( \frac{t^{2\alpha+1}}{2\alpha+1}\right)+a_1\right\rbrace ^{\frac{\beta}{3}-\omega-1}
\end{equation}
where $a_2$ is an arbitrary constant.\\
Then from equations (\ref{eqn:16}, \ref{eqn:19}) we have
\begin{equation}
\label{eqn:20}
p_d=\frac{\omega a_2}{G}(\gamma t^{-\alpha})^{-3\omega-1}\left\lbrace 3a_0\gamma^{-2}\left( \frac{t^{2\alpha+1}}{2\alpha+1}\right)+a_1\right\rbrace ^{\frac{\beta}{3}-\omega-1}
\end{equation}
Now equations (\ref{eqn:9}, \ref{eqn:19}) gives
\begin{multline}
\rho_m=\frac{1}{8\pi G}\biggl[3a_0^2\gamma^{-4}t^{4\alpha}\left\lbrace 3a_0\gamma^{-2}\left( \frac{t^{2\alpha+1}}{2\alpha+1}\right)+a_1\right\rbrace^{-2}-
\alpha (\alpha+1)t^{-2}+ 3\alpha^2 t^{-2}-3\alpha a_0 \gamma^{-2}t^{2\alpha-1} \\
 \times \left\lbrace 3a_0\gamma^{-2}\left( \frac{t^{2\alpha+1}}{2\alpha+1}\right)+a_1\right\rbrace^{-1}-8\pi a_2(\gamma t^{-\alpha})^{-3\omega-1}\left\lbrace 3a_0\gamma^{-2}\left( \frac{t^{2\alpha+1}}{2\alpha+1}\right)+a_1\right\rbrace ^{\frac{\beta}{3}-\omega-1}\biggr]
\label{eqn:21}
\end{multline}\\
Again subtracting $3$ times equation (\ref{eqn:10}) from $2$ times equation (\ref{eqn:11}) we have
\begin{align*}
8\pi G p_m=3H^2+3H\frac{\dot{\phi}}{\phi}-\frac{\ddot{\phi}}{\phi}+\biggl(\frac{\dot{\phi}}{\phi}\biggr)^2-8\pi G p_d
\end{align*}
which gives
\begin{multline}
\label{eqn:22}
p_m=\frac{1}{8\pi G}\biggl[3a_0^2\gamma^{-4}t^{4\alpha}\left\lbrace 3a_0\gamma^{-2}\left( \frac{t^{2\alpha+1}}{2\alpha+1}\right)+a_1\right\rbrace^{-2}
-8\pi\omega a_2(\gamma t^{-\alpha})^{-3\omega-1}\left\lbrace 3a_0\gamma^{-2}\left( \frac{t^{2\alpha+1}}{2\alpha+1}\right)+a_1\right\rbrace ^{\frac{\beta}{3}-\omega-1} \\
-\alpha t^{-2}-3\alpha a_0 \gamma^{-2}t^{2\alpha-1} \left\lbrace 3a_0\gamma^{-2}\left( \frac{t^{2\alpha+1}}{2\alpha+1}\right)+a_1\right\rbrace^{-1} \biggr]
\end{multline}
And here
\begin{equation}
\label{eqn:23}
 Q=\beta a_0 a_2 G^{-1}\gamma^{-3(1+\omega)}t^{3\alpha+3\alpha\omega} \left\lbrace 3a_0\gamma^{-2}\left( \frac{t^{2\alpha+1}}{2\alpha+1}\right)+a_1\right\rbrace^{\frac{\beta}{3}-\omega-2} 
\end{equation}
The behavior of energy density, pressure and interaction term are shown as a function of cosmic time in Figures [\ref{fig1}, \ref{fig2}, \ref{fig2a}].

\begin{figure}[H]
\centering
\includegraphics[scale=0.4]{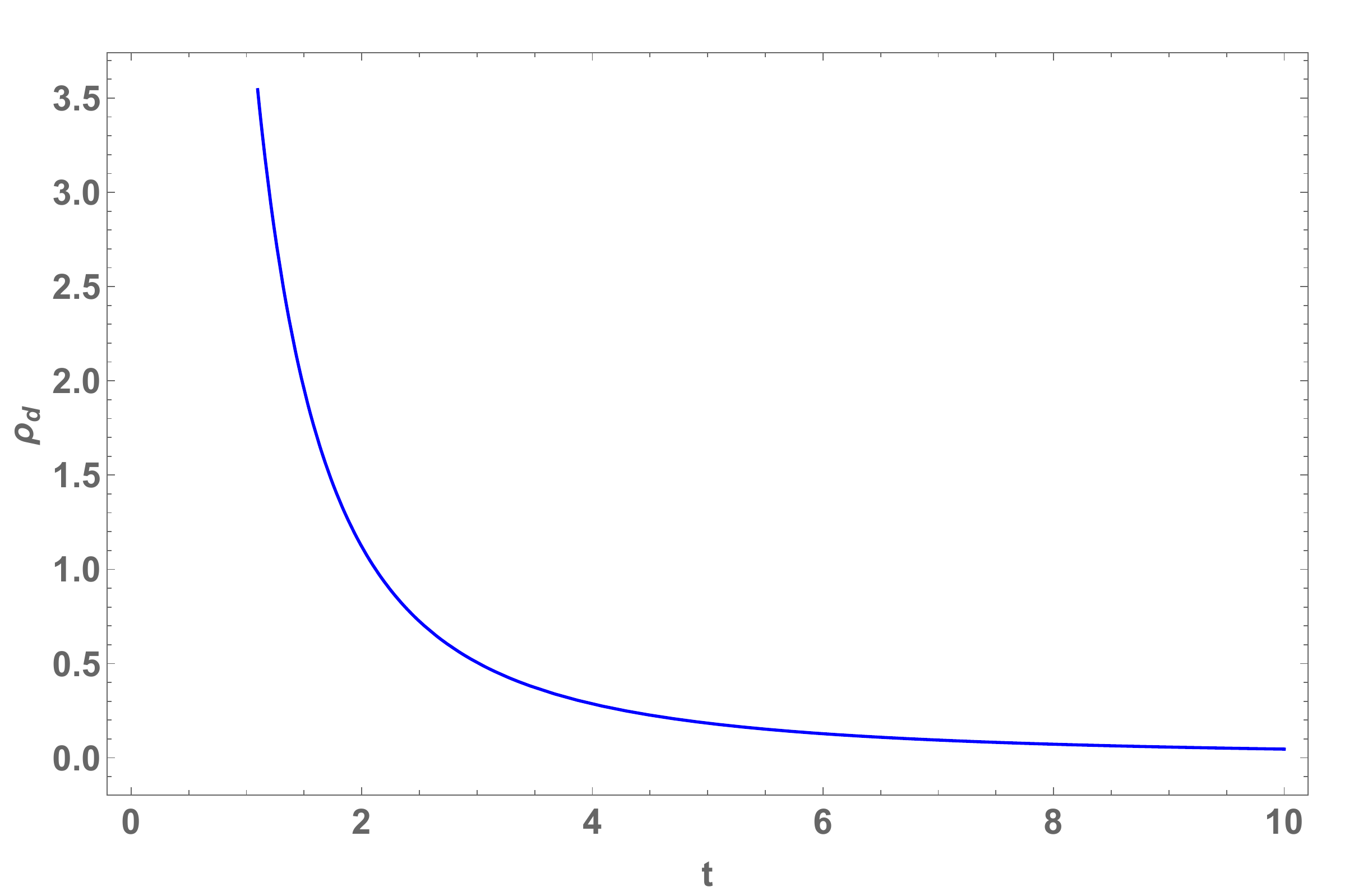}
\caption{Energy density `$\rho_d$' vs time `$t$'(in Gyr) for $\omega=-0.95$, $\beta=0.01$, $\gamma=2$, $a_0=1.5$, $a_1=1.3$, $a_2=1.2$, $G=\alpha=1$.}
\label{fig1}
\end{figure}

\begin{figure}[H]
\centering
\includegraphics[scale=0.4]{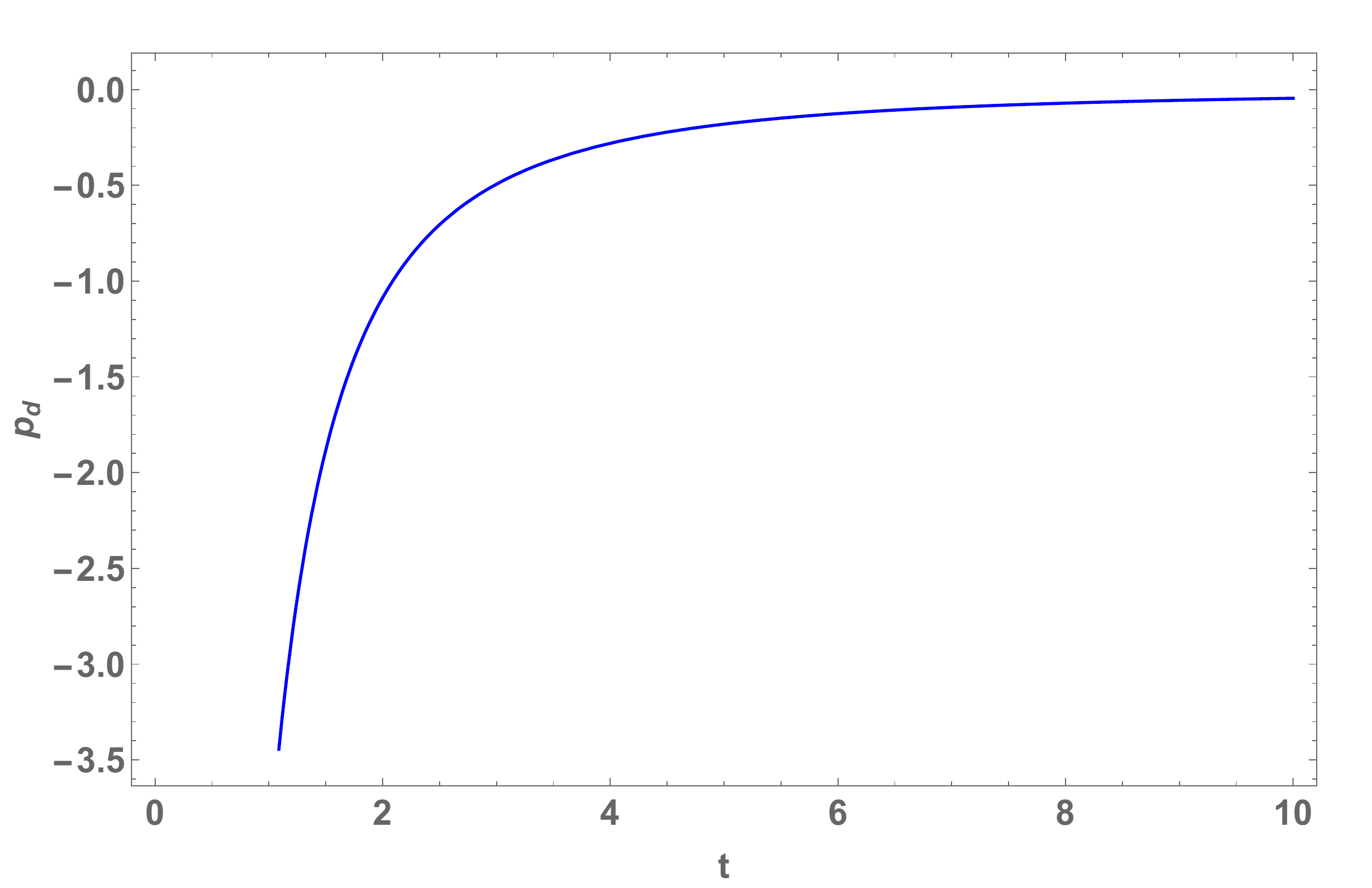}
\caption{Pressure `$p_d$' of dark energy vs time `$t$'(in Gyr) for $\omega=-0.95$, $\beta=0.01$, $\gamma=2$, $a_0=1.5$, $a_1=1.3$, $a_2=1.2$, $G=\alpha=1$.}
\label{fig2}
\end{figure}

\begin{figure}[H]
\centering
\includegraphics[scale=0.4]{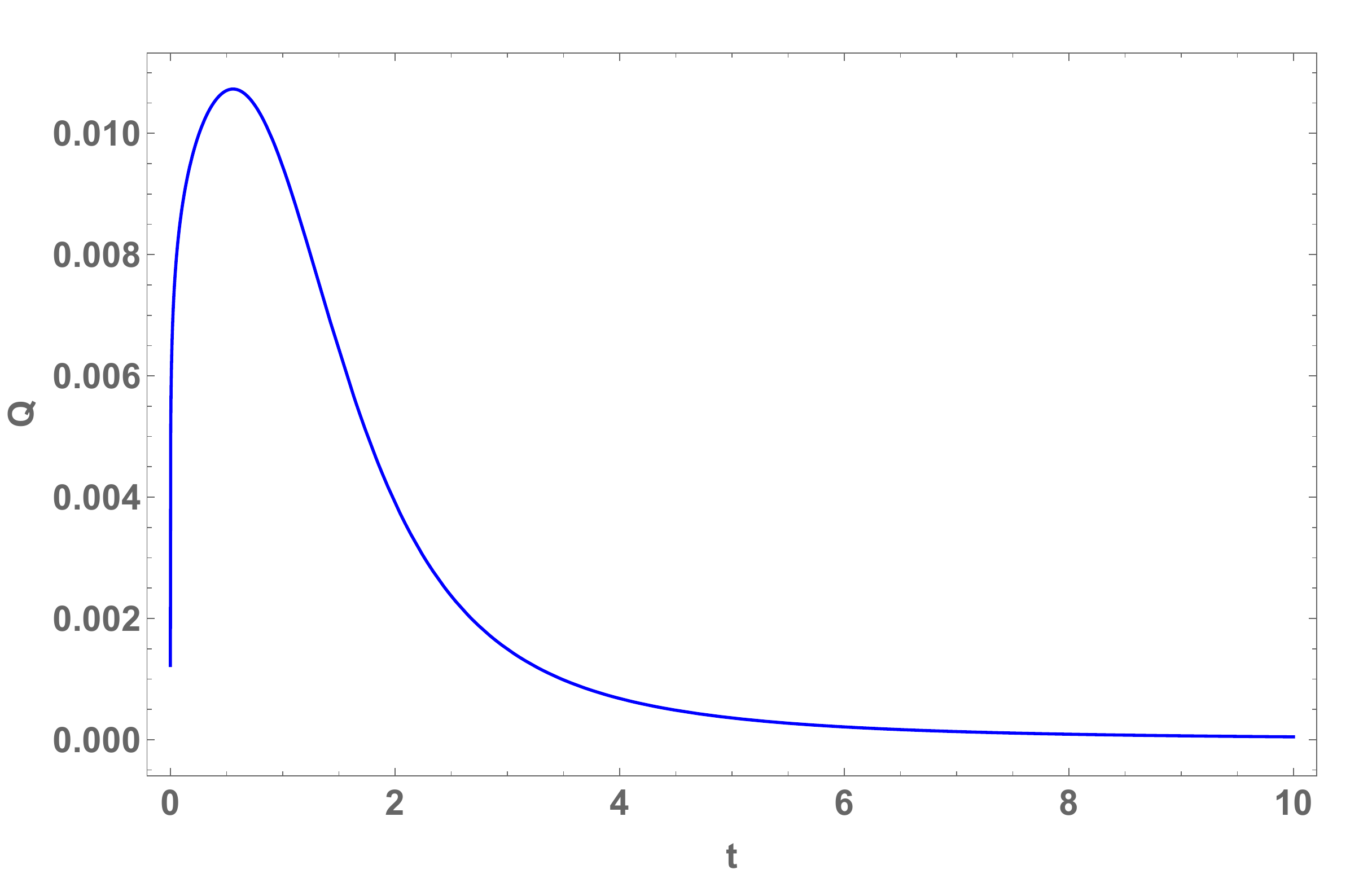}
\caption{Interaction term `$Q$' of dark energy vs time `$t$'(in Gyr) for $\omega=-0.95$, $\beta=0.01$, $\gamma=2$, $a_0=1.5$, $a_1=1.3$, $a_2=1.2$, $G=\alpha=1$.}
\label{fig2a}
\end{figure}
\textbf{Case II :}

In this case we take the interaction term $Q$ to be
\begin{equation}
\label{eqn:24}
Q=\beta H\rho_m
\end{equation}
Then equations (\ref{eqn:13}) and (\ref{eqn:14}) respectively take the forms
\begin{equation}
\label{eqn:25}
\dot{\rho_d}+(\rho_d+p_d)3H+\rho_d\biggl(\frac{\dot{G}}{G}+\frac{\dot{\phi}}{\phi}\biggr)+3p_d\biggl(\frac{\dot{\phi}}{\phi}\biggr)=\beta H\rho_m
\end{equation}
and
\begin{equation}
\label{eqn:26}
\dot{\rho_m}+(\rho_m+p_m)3H+\rho_m\biggl(\frac{\dot{G}}{G}+\frac{\dot{\phi}}{\phi}\biggr)+3p_m\biggl(\frac{\dot{\phi}}{\phi}\biggr)=-\beta H\rho_m
\end{equation}
Now for the cases $\alpha=\frac{1}{2}, \pm 1$ and $\alpha\sim\frac{-1}{2}$,
 we have, as usual, from equation (\ref{eqn:15}),
\begin{equation}
\label{eqn:27}
a(t)=\left\lbrace 3a_0\gamma^{-2}\left(\frac{t^{2\alpha+1}}{2\alpha+1}\right)\right\rbrace^{\frac{1}{3}}
\end{equation}
where $a_0$ is an arbitrary constant.\\
Then equation (\ref{eqn:9}) gives
\begin{equation}
\label{eqn:28}
8\pi G\rho_m=\frac{1}{3}(4\alpha^2-2\alpha+1)t^{-2}-8\pi G\rho_d
\end{equation}
Thus now from equations (\ref{eqn:25}) and (\ref{eqn:28}) we have
\begin{equation}
\label{eqn:29}
\dot{\rho_d}+(\rho_d+p_d)3H+\rho_d\biggl(\frac{\dot{G}}{G}+\frac{\dot{\phi}}{\phi}\biggr)+3p_d\biggl(\frac{\dot{\phi}}{\phi}\biggr)=\frac{\beta H}{8\pi G  }\biggl[\frac{1}{3}(4\alpha^2-2\alpha+1)t^{-2}-8\pi G\rho_d\biggr]
\end{equation}
which gives
\begin{equation}
\label{eqn:30}
\rho_d=\frac{\beta(1+2\alpha)}{72\pi G}\biggl[\frac{4\alpha^2-2\alpha+1}{\alpha-\alpha\omega+\frac{\beta}{3}(2\alpha+1)+\omega-1}\biggr]t^{-2}
\end{equation}
Now using equation (\ref{eqn:30}) we get
\begin{equation}
\label{eqn:31}
p_d=\frac{\omega\beta(1+2\alpha)}{72\pi G}\biggl[\frac{4\alpha^2-2\alpha+1}{\alpha-\alpha\omega+\frac{\beta}{3}(2\alpha+1)+\omega-1}\biggr]t^{-2}
\end{equation}
And from equations (\ref{eqn:9}) and (\ref{eqn:30}) we have
\begin{equation}
\label{eqn:32}
\rho_m=\frac{1}{24\pi G}(4\alpha^2-2\alpha+1)t^{-2}-\frac{\beta(1+2\alpha)}{72\pi G}\biggl[\frac{4\alpha^2-2\alpha+1}{\alpha-\alpha\omega+\frac{\beta}{3}(2\alpha+1)+\omega-1}\biggr]t^{-2}
\end{equation}
Again from equations (\ref{eqn:11}, \ref{eqn:31}) we get
\begin{equation}
\label{eqn:33}
p_m=-\frac{1}{24\pi G}(\alpha^2+2\alpha-1)t^{-2}-\frac{\omega\beta(1+2\alpha)}{72\pi G}\biggl[\frac{4\alpha^2-2\alpha+1}{\alpha-\alpha\omega+\frac{\beta}{3}(2\alpha+1)+\omega-1}\biggr]t^{-2}
\end{equation}
And here
\begin{equation}
\label{eqn:34}
Q=\frac{\beta(1+2\alpha)}{72\pi G}(4\alpha^2-2\alpha+1)t^{-3}-\frac{\beta^2(1+2\alpha)^2}{216\pi G}\biggl[\frac{4\alpha^2-2\alpha+1}{\alpha-\alpha\omega+\frac{\beta}{3}(2\alpha+1)+\omega-1}\biggr]t^{-3}
\end{equation}

The behavior of energy density, pressure and interaction term are shown as a function of cosmic time in Figures [\ref{fig3}, \ref{fig4}, \ref{fig4a}].

\begin{figure}[H]
\centering
\includegraphics[scale=0.4]{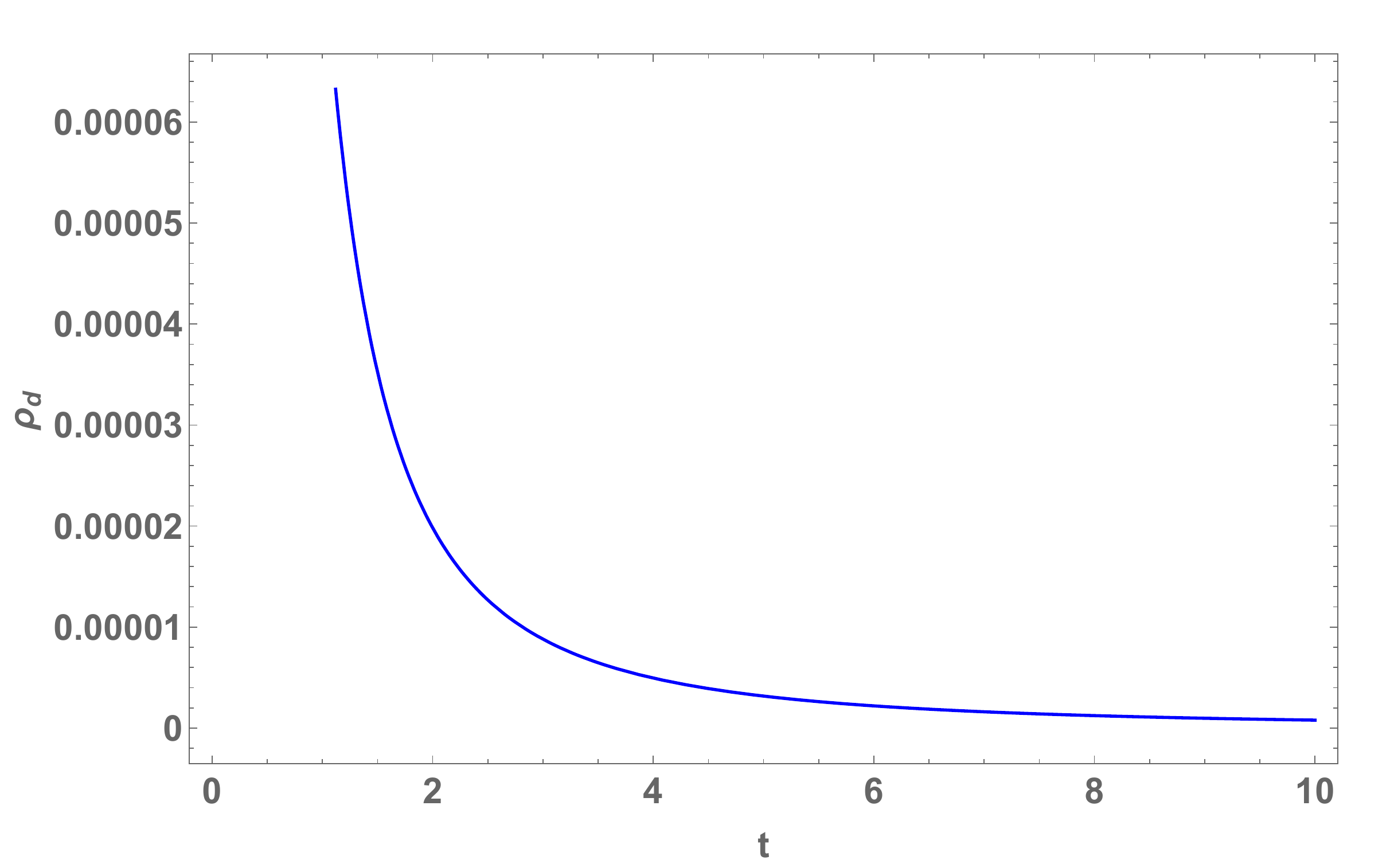}
\caption{Energy density `$\rho_d$' vs time `$t$'(in Gyr) for $\omega=-0.95$, $\beta=0.01$, $\alpha=-1$, $G=1$.}
\label{fig3}
\end{figure}

\begin{figure}[H]
\centering
\includegraphics[scale=0.4]{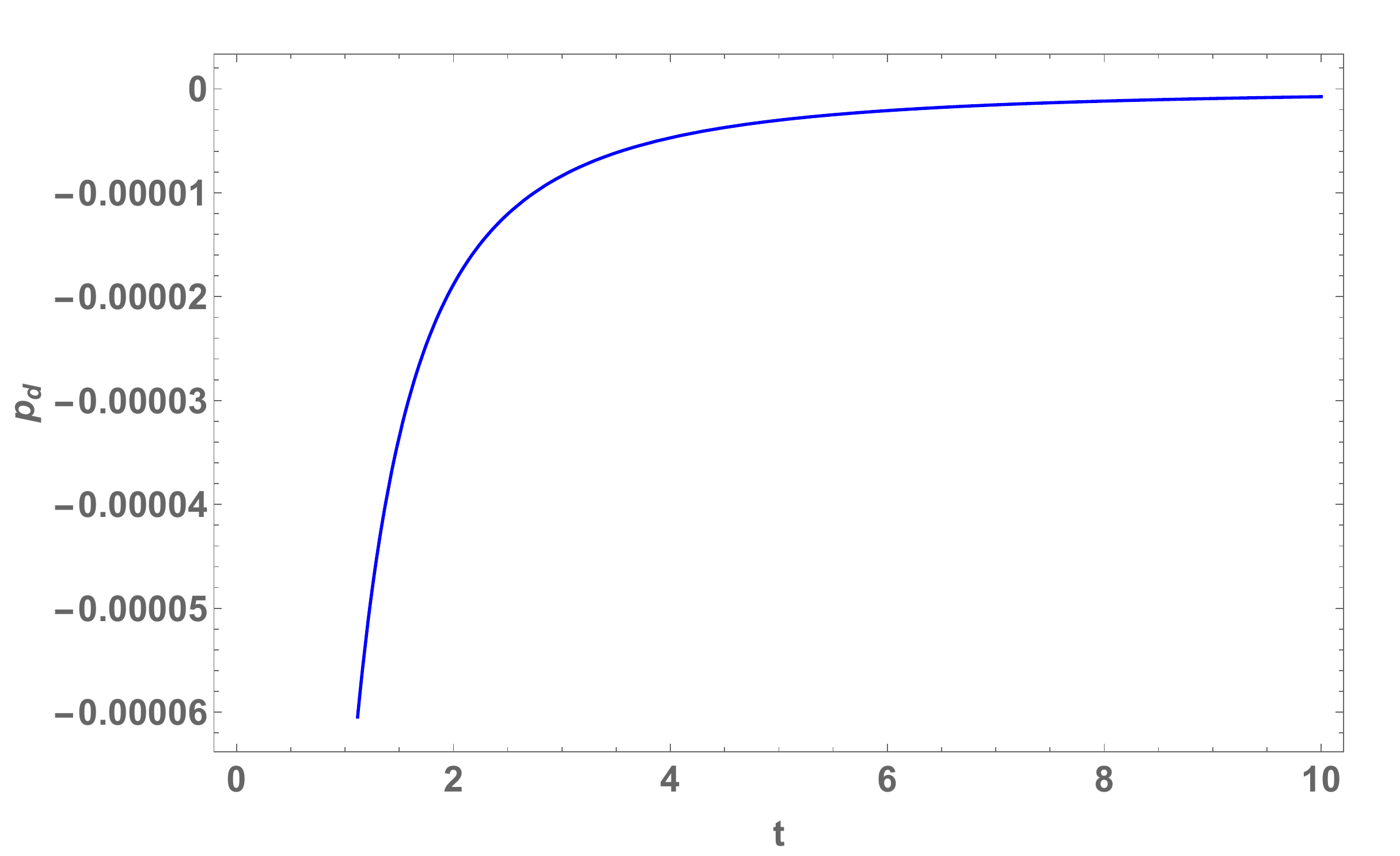}
\caption{Pressure `$p_d$' of dark energy vs time `$t$'(in Gyr) for $\omega=-0.95$, $\beta=0.01$, $\alpha=-1$, $G=1$.}
\label{fig4}
\end{figure}

\begin{figure}[H]
\centering
\includegraphics[scale=0.4]{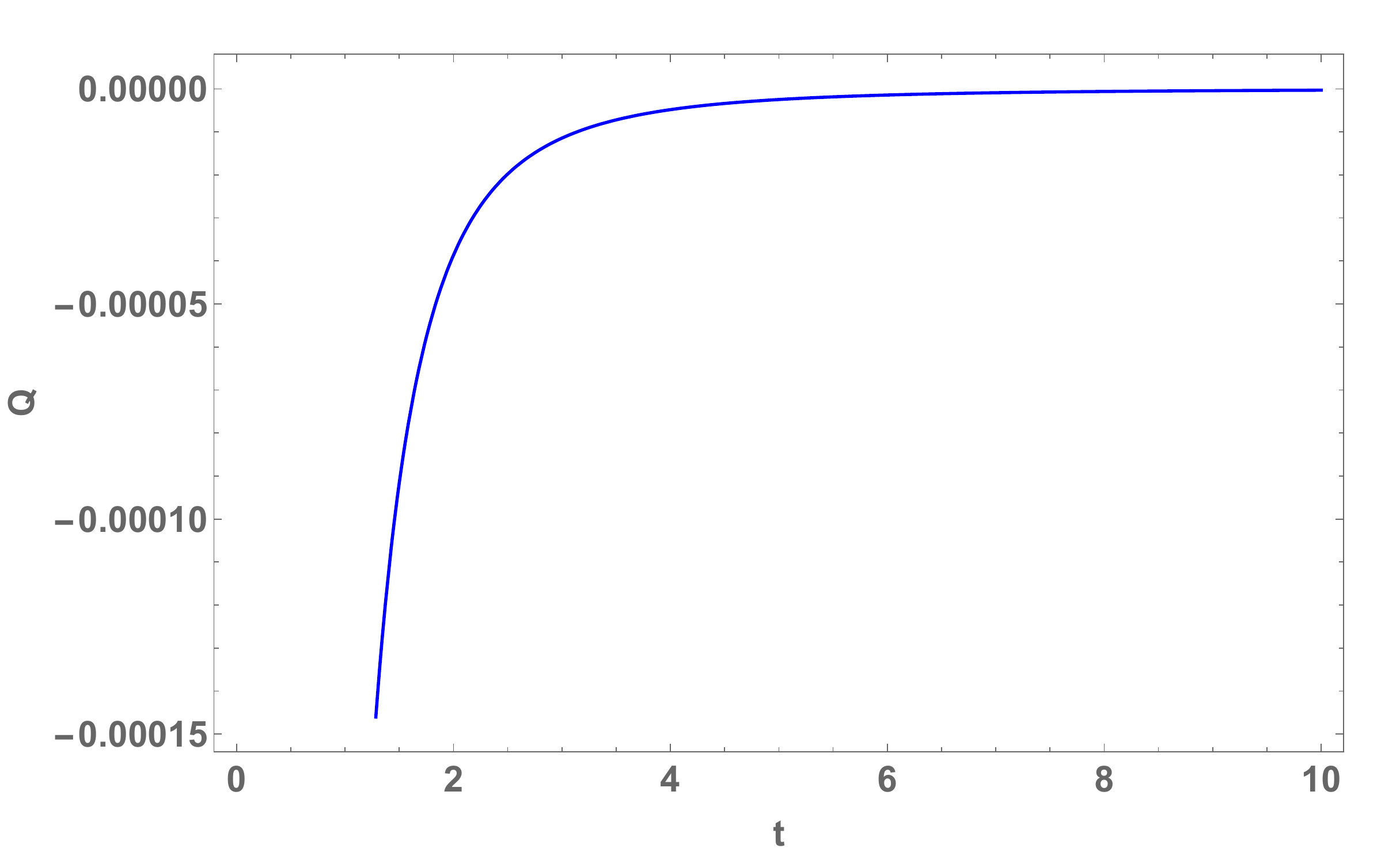}
\caption{Interaction term `$Q$' of dark energy vs time `$t$'(in Gyr) for $\omega=-0.95$, $\beta=0.01$, $\alpha=-1$, $G=1$.}
\label{fig4a}
\end{figure}

\textbf{Case III :}
Here we take up the case of relation (5) when
\begin{equation}
\label{eqn:35}
\phi=t^\frac{1}{2}
\end{equation}
Then from (15) we get
\begin{equation}
\label{eqn:36}
a(t)=\left\lbrace  \log(b_1 t^b)\right\rbrace^{\frac{1}{3}},
\end{equation}
where $b$, $b_1$ are constants.
Here we take, as in case I,
\begin{align*}
Q=\beta H \rho_d
\end{align*}
Then from equation (\ref{eqn:13}) we get
\begin{equation}
\label{eqn:37}
\rho_d=b_2G^{-1} t^{-\frac{1}{2}(1+3\omega)}\lbrace\log(b_1 t^b)\rbrace^{\frac{\beta}{3}-\omega-1}
\end{equation}
 $b_2$ being an constant. And in this case
\begin{equation}
\label{eqn:38}
p_d=\omega b_2G^{-1} t^{-\frac{1}{2}(1+3\omega)}\lbrace\log(b_1 t^b)\rbrace^{\frac{\beta}{3}-\omega-1}
\end{equation}
Thus equation (\ref{eqn:9}) gives 
\begin{equation}
\label{eqn:39}
\rho_m=\frac{1}{8\pi G}\biggl[ t^{-2}+\frac{b^2}{3}t^{-2}\{\log(b_1 t^b)\}^{-2}
+\frac{1}{2}bt^{-2}{\log(b_1 t^b)}^{-1}
-8\pi b_2 t^{-\frac{1}{2}(1+3\omega)}\lbrace\log(b_1 t^b)\rbrace^{\frac{\beta}{3}-\omega-1}\biggr]
\end{equation}
Again from equations (\ref{eqn:10}), (\ref{eqn:11}) we have
\begin{equation}
\label{eqn:40}
p_m=\frac{1}{8\pi G}\biggl[\frac{1}{2} t^{-2}+\frac{b^2}{3}t^{-2}\{\log(b_1 t^b)\}^{-2}
+\frac{1}{2}bt^{-2}{\log(b_1 t^b)}^{-1}
-8\pi\omega b_2 t^{-\frac{1}{2}(1+3\omega)}\lbrace\log(b_1 t^b)\rbrace^{\frac{\beta}{3}-\omega-1}\biggr]
\end{equation}
Here,
\begin{equation}
\label{eqn:41}
Q= \frac{\beta}{3}b b_2G^{-1} t^{-\frac{1}{2}(3+3\omega)}\lbrace\log(b_1 t^b)\rbrace^{\frac{\beta}{3}-\omega-2}
\end{equation}
And, squared velocity of sound
\begin{equation}
\label{eqn:42}
v_s^2=\frac{\delta p}{\delta\rho}>0
\end{equation}

The behavior of energy density, pressure and interaction term are shown as a function of cosmic time in Figures [\ref{fig5}, \ref{fig6}, \ref{fig6a}].

\begin{figure}[H]
\centering
\includegraphics[scale=0.4]{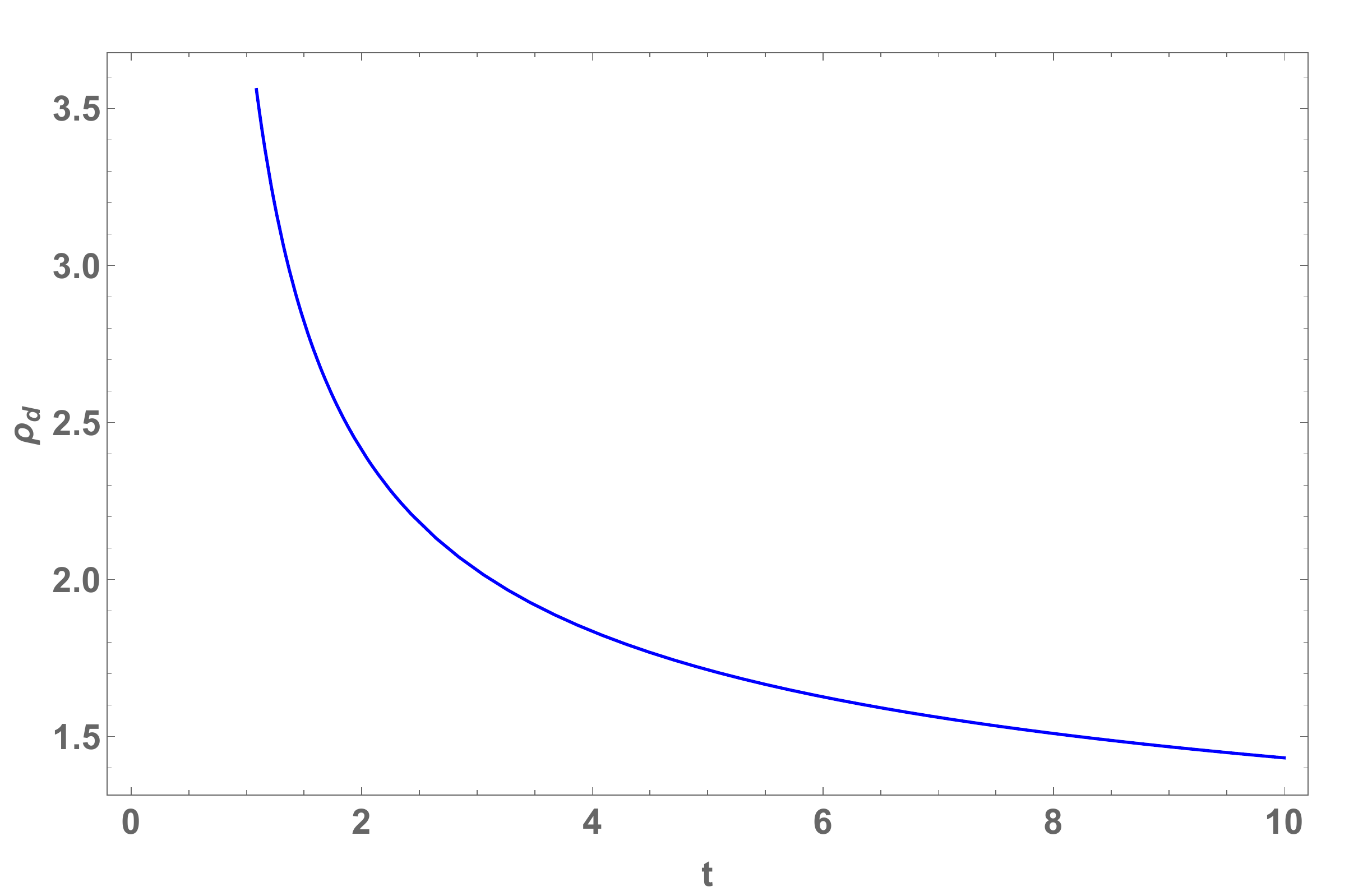}
\caption{Energy density `$\rho_d$' vs time `$t$'(in Gyr) for $\omega=-0.33$, $\beta=0.01$, $b=1.2$, $b_1=2.3$, $b_2=3.4$, $G=1$.}
\label{fig5}
\end{figure}

\begin{figure}[H]
\centering
\includegraphics[scale=0.4]{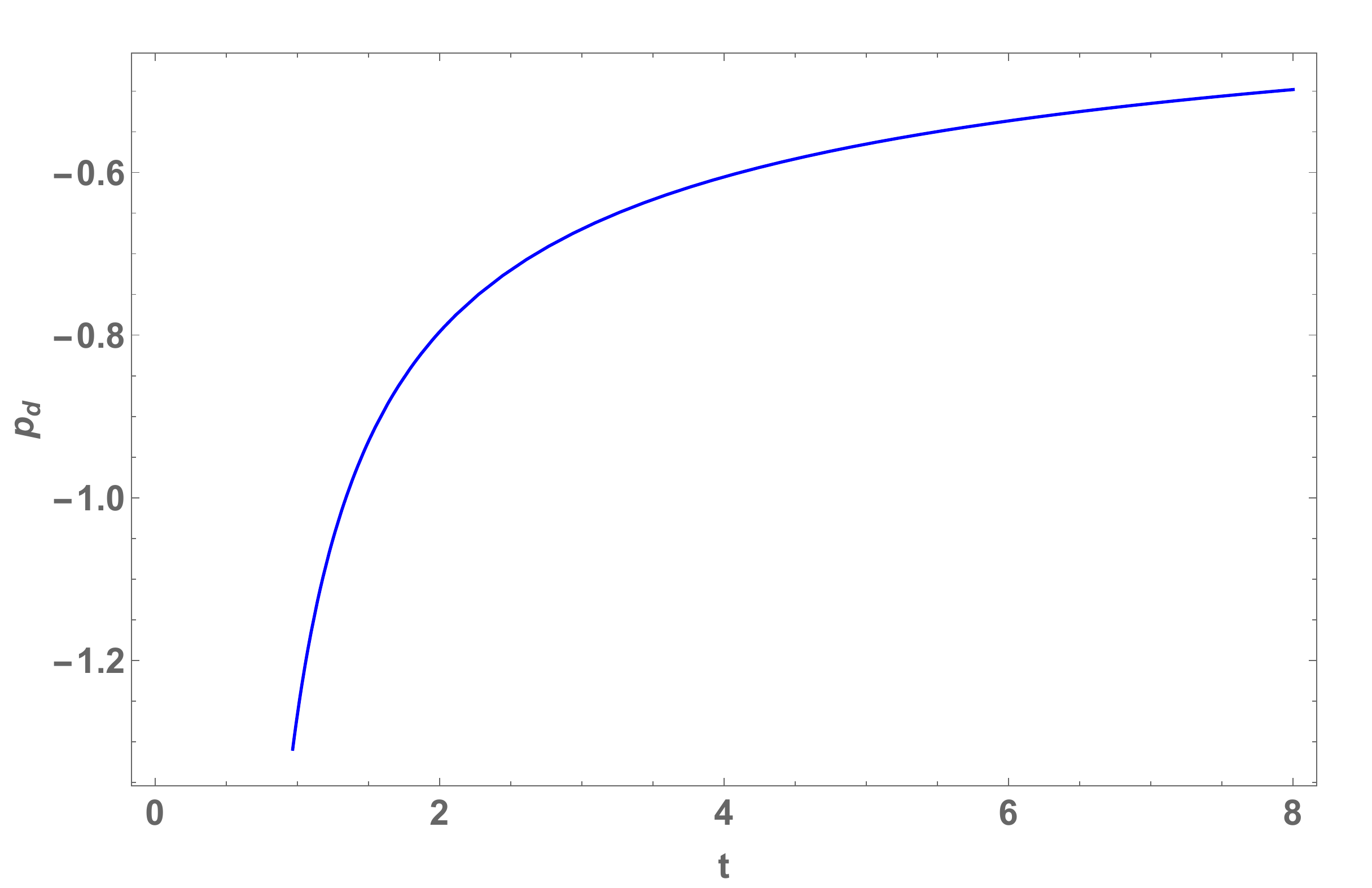}
\caption{Pressure `$p_d$' of dark energy vs time `$t$'(in Gyr) for $\omega=-0.33$, $\beta=0.01$, $b=1.2$, $b_1=2.3$, $b_2=3.4$, $G=1$.}
\label{fig6}
\end{figure}
\begin{figure}[H]
\centering
\includegraphics[scale=0.4]{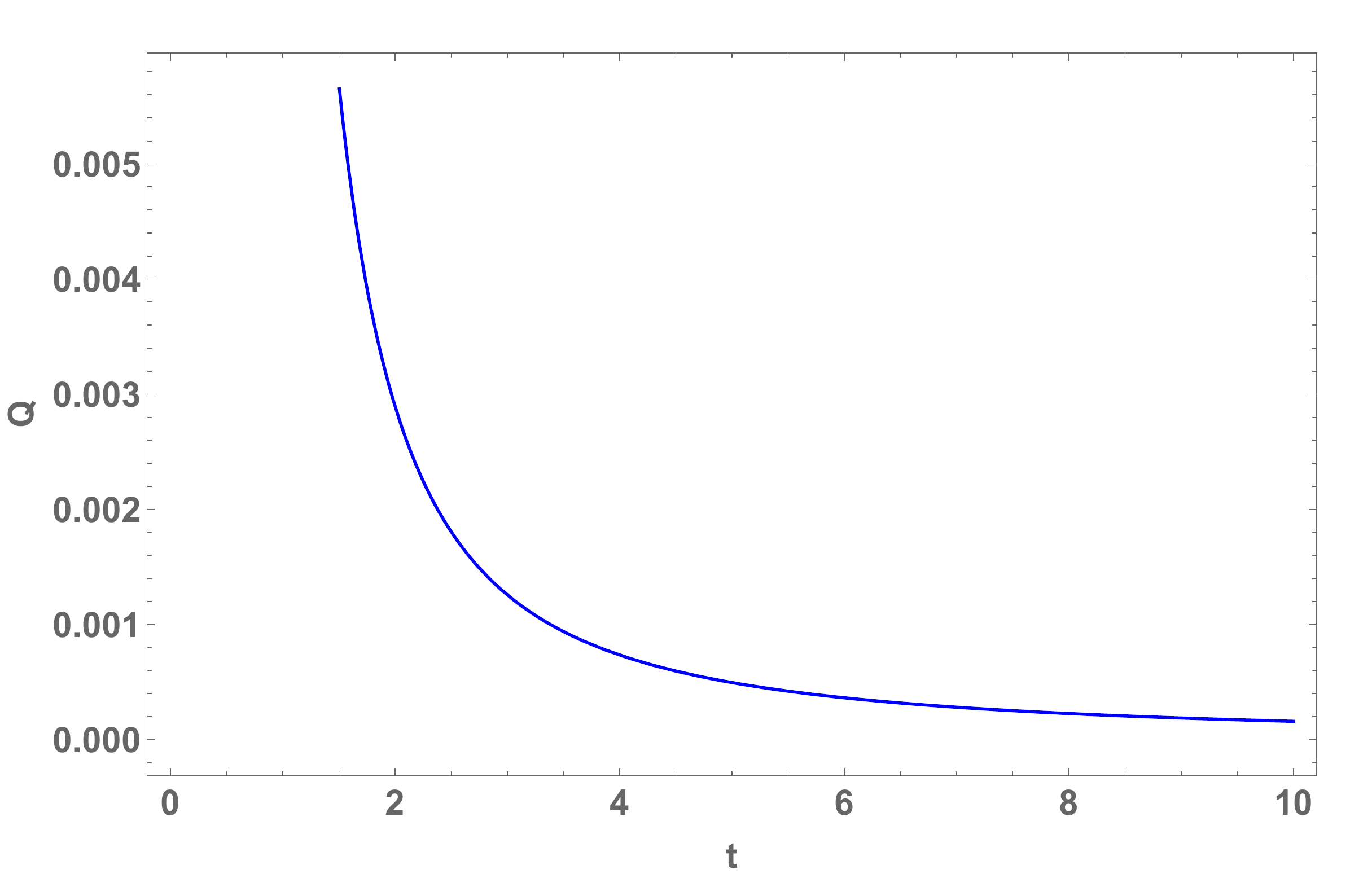}
\caption{Interaction term `$Q$' of dark energy vs time `$t$'(in Gyr) for $\omega=-0.33$, $\beta=0.01$, $b=1.2$, $b_1=2.3$, $b_2=3.4$, $G=1$.}
\label{fig6a}
\end{figure}

\textbf{Case IV :}
In this case we take
\begin{equation}
\label{eqn:43}
\phi=t^{\frac{1}{2}}
\end{equation}
and consider the interaction $Q$ as
\begin{equation}
\label{eqn:44}
Q=\beta H\rho_m
\end{equation}
Now from equations  (\ref{eqn:9}), (\ref{eqn:15}), (\ref{eqn:43}) we have
\begin{equation}
\label{eqn:45}
8\pi G\rho_m=\frac{1}{3}c_2^2t^{-2}(\log (c_1 t^{c_2}))^{-2}
+\frac{c_2}{2}t^{-2}(\log (c_1 t^{c_2}))^{-1}+t^{-2}-8\pi G\rho_d
\end{equation}
where $c_1$ and $c_2$ are arbitrary constants.\\
Then equations (\ref{eqn:13}, \ref{eqn:45}) give
\begin{equation}
\label{eqn:46}
\dot{\rho_d}+(\rho_d+\omega\rho_d)3H+\rho_d \biggl(\frac{\dot{G}}{G}+\frac{\dot{\phi}}{\phi}\biggr)
+3\omega\rho_d\frac{\dot{\phi}}{\phi}
=\frac{\beta H}{8\pi G} \biggl[ \frac{1}{3}c_2^2t^{-2}(\log (c_1 t^{c_2}))^{-2}
+\frac{c_2}{2}t^{-2}(\log (c_1 t^{c_2}))^{-1}+t^{-2}-8\pi G\rho_d\biggr]
\end{equation}
Now one of the solutions of this equation is 
\begin{equation}
\label{eqn:47}
\rho_d=G^{-1}t^{-2}\biggl[ \frac{c_2^2\beta}{24\pi(\beta-4)}(\log (c_1 t^{c_2}))^{-2}
+\frac{3c_2\beta(\beta-2)}{48\pi(\beta-1)(\beta-4)}(\log (c_1t^{c_2}))^{-1}\biggr]
\end{equation}
with $\omega=\frac{-1}{3}$.
Using equation (\ref{eqn:47}) we get
\begin{equation}
\label{eqn:48}
p_d=\frac{-1}{3}G^{-1}t^{-2}\biggl[ \frac{c_2^2\beta}{24\pi(\beta-4)}(\log (c_1 t^{c_2}))^{-2}
+\frac{3c_2\beta(\beta-2)}{48\pi(\beta-1)(\beta-4)}(\log (c_1t^{c_2}))^{-1}\biggr]
\end{equation}
Now from (\ref{eqn:9}) we get
\begin{multline}
\label{eqn:49}
\rho_m=\frac{1}{8\pi G}\biggl[\frac{c_2^2}{3}(\log (c_1t^{c_2}))^{-2}+\frac{{c_2}}{2}(\log (c_1 t^{c_2}))^{-1}+1 \biggr]t^{-2}
-G^{-1}t^{-2}\biggl[ \frac{c_2^2\beta}{24\pi(\beta-4)}(\log (c_1 t^{c_2}))^{-2}\\
+\frac{3c_2\beta(\beta-2)}{48\pi(\beta-1)(\beta-4)}(\log (c_1t^{c_2}))^{-1}\biggr]
\end{multline}
And equation (\ref{eqn:11}) gives 
\begin{multline}
\label{eqn:50}
p_m=\frac{1}{8\pi G}\biggl[\frac{1}{3}c_2^2(\log (c_1 t^{c_2}))^{-2}+\frac{c_2}{2}(\log (c_1t^{c_2}))^{-1}+\frac{1}{2}\biggr]t^{-2}
-\frac{1}{3}G^{-1}t^{-2}\biggl[ \frac{c_2^2\beta}{24\pi(\beta-4)}(\log (c_1 t^{c_2}))^{-2}\\
+\frac{3c_2\beta(\beta-2)}{48\pi(\beta-1)(\beta-4)}(\log (c_1t^{c_2}))^{-1}\biggr]
\end{multline}
The interaction,
\begin{multline}
\label{eqn:51}
Q=\frac{c_2\beta}{24\pi G}\biggl[\frac{c_2^2}{3}(\log (c_1t^{c_2}))^{-3}+\frac{c_2}{2}(\log (c_1 t^{c_2}))^{-2}+(\log (c_1 t^{c_2}))^{-1} \biggr]t^{-3}
-\frac{c_2\beta}{3G}\biggl[ \frac{c_2^2\beta}{24\pi(\beta-4)}(\log (c_1 t^{c_2}))^{-3}\\
+\frac{3c_2\beta(\beta-2)}{48\pi(\beta-1)(\beta-4)}(\log (c_1t^{c_2}))^{-2}\biggr] t^{-3}
\end{multline}

The behavior of energy density and pressure are shown as a function of cosmic time in Figures [\ref{fig7}-\ref{fig8}].

\begin{figure}[H]
\centering
\includegraphics[scale=0.4]{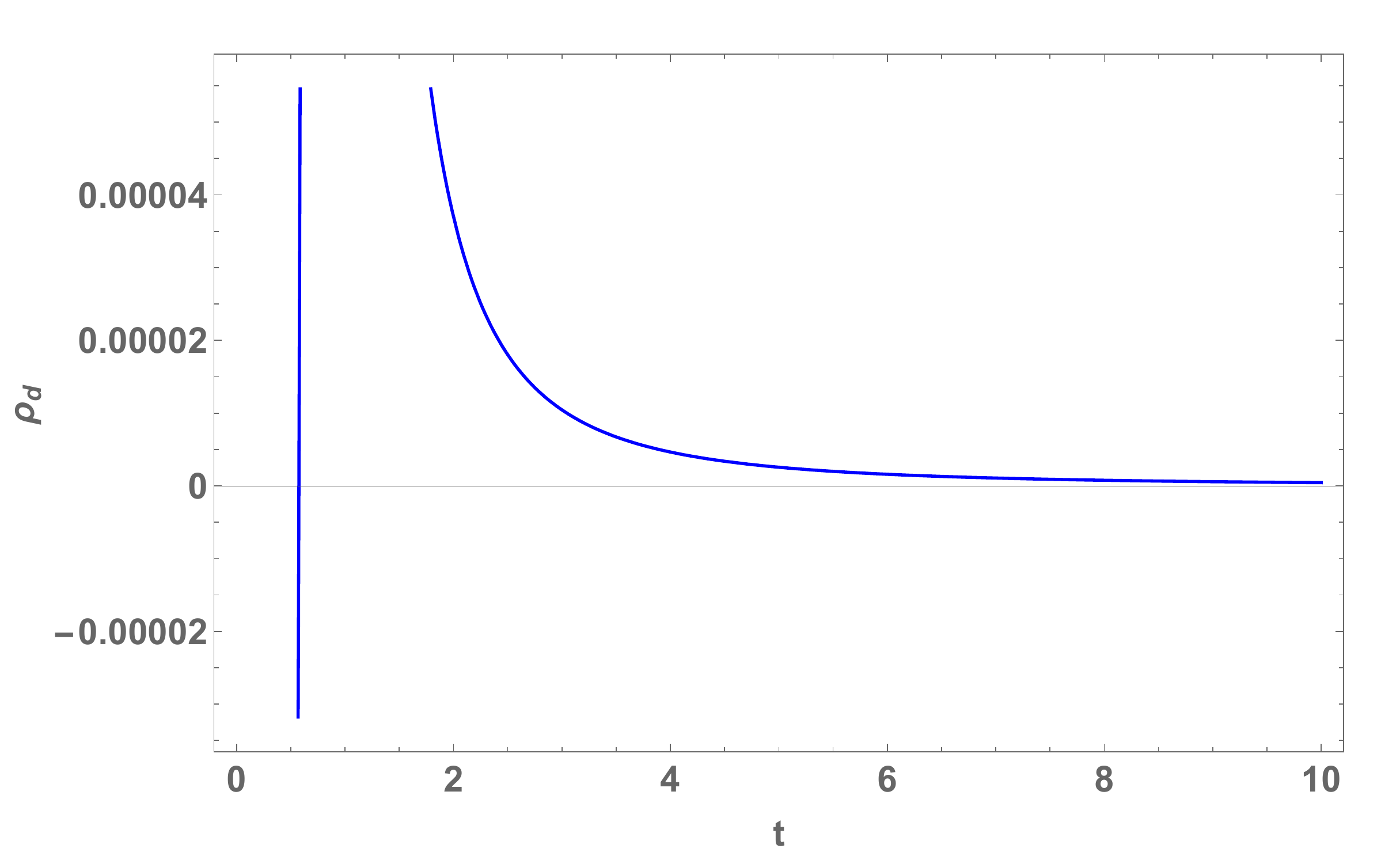}
\caption{Energy density `$\rho_d$' vs time `$t$'(in Gyr) for $\omega=-0.33$, $\beta=-0.01$, $c_1=1.3$, $c_2=1.2$, $G=1$.}
\label{fig7}
\end{figure}

\begin{figure}[H]
\centering
\includegraphics[scale=0.4]{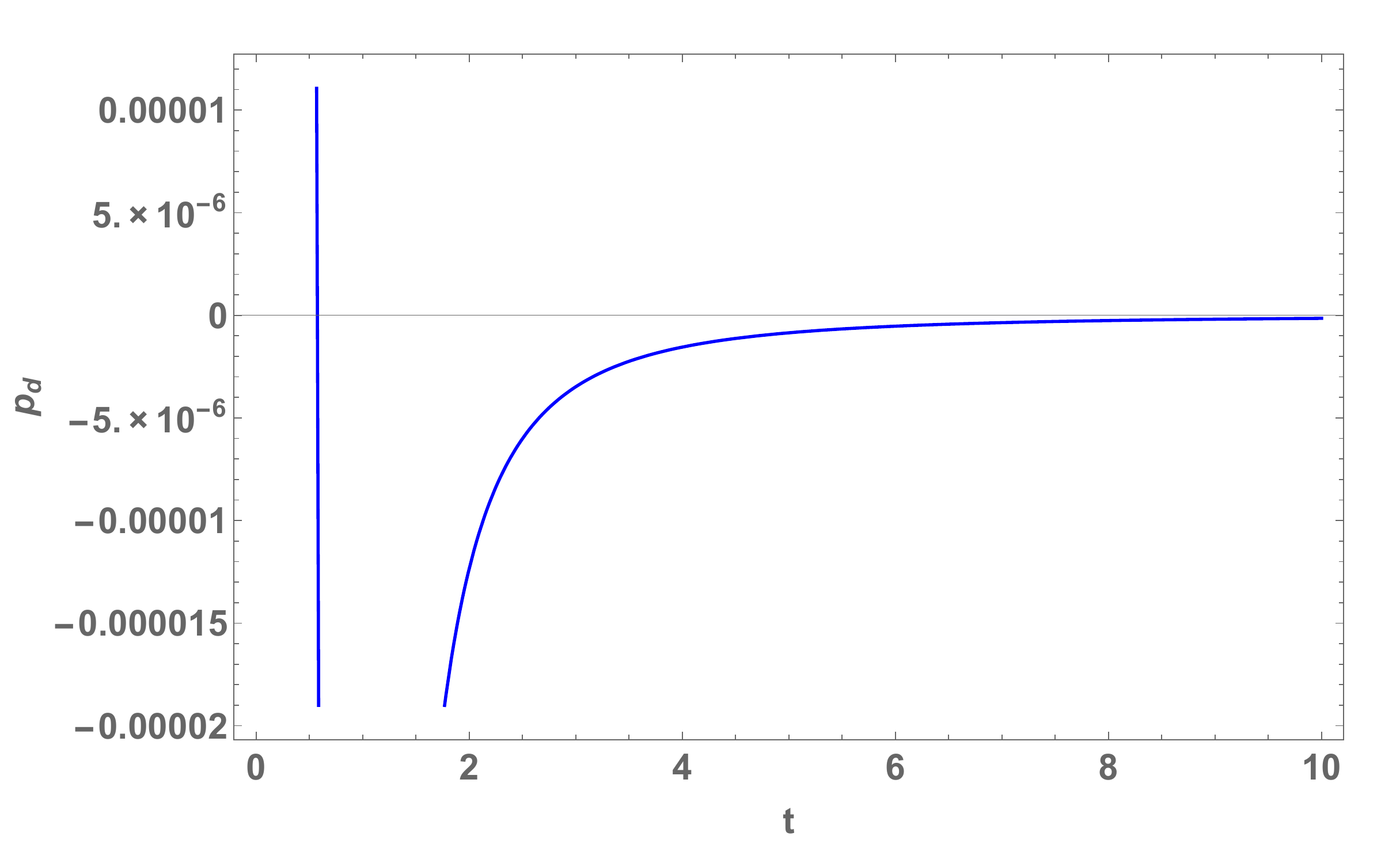}
\caption{Pressure `$p_d$' of dark energy vs time `$t$'(in Gyr)for $\omega=-0.33$, $\beta=-0.01$, $c_1=1.3$, $c_2=1.2$, $G=1$.}
\label{fig8}
\end{figure}
\begin{figure}[H]
\centering
\includegraphics[scale=0.4]{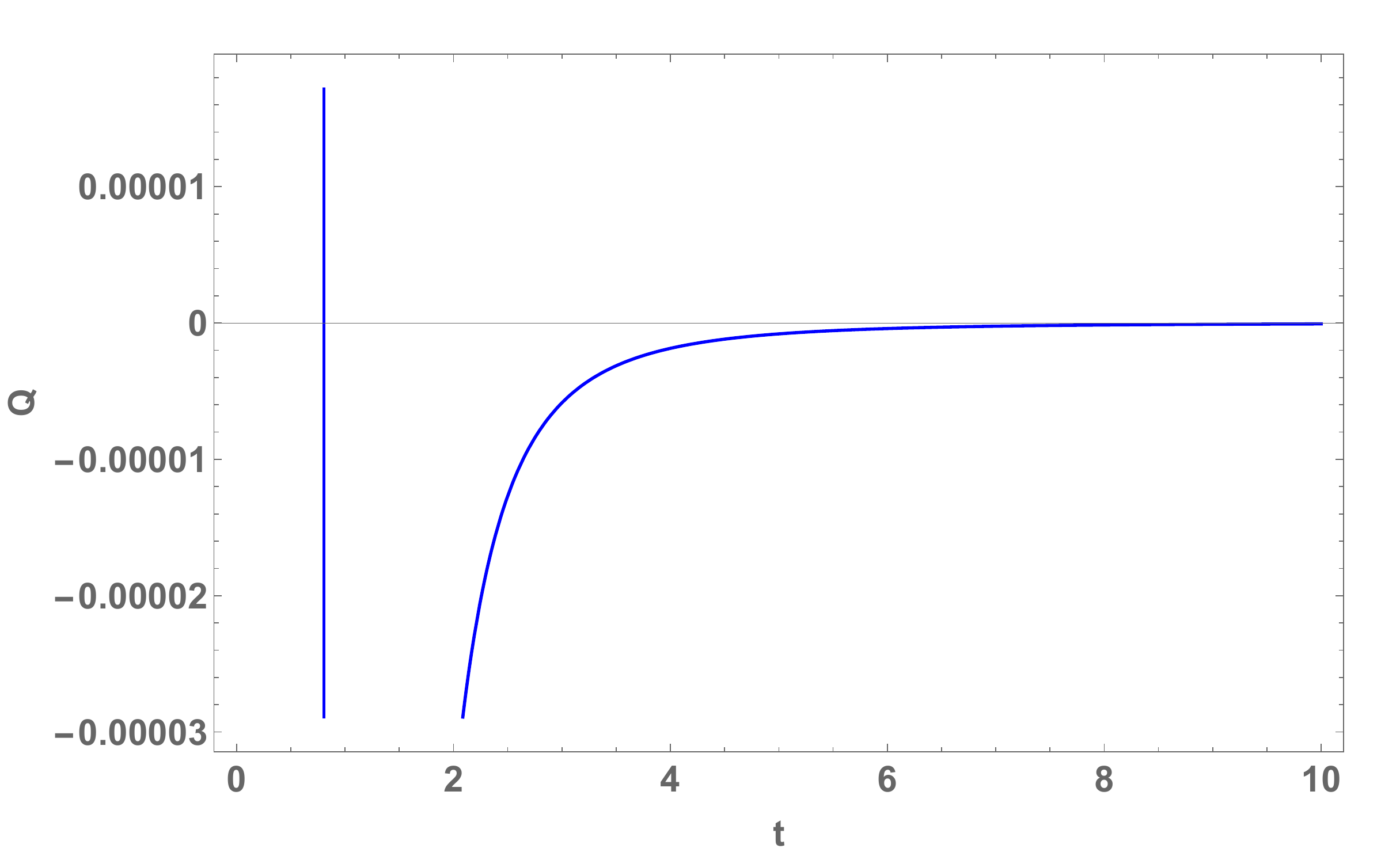}
\caption{Interaction term `$Q$' of dark energy vs time `$t$'(in Gyr) for for $\omega=-0.33$, $\beta=-0.01$, $c_1=1.3$, $c_2=1.2$, $G=1$.}
\label{fig8a}
\end{figure}

\section{Study of the results and conclusions obtained}
In the universe obtained in case I we see that the interaction between dark energy and other components of the universe is occurring almost at the beginning of the evolution. Here the constants $\gamma, a_0$ and $a_1$ seem to play significant roles in the history of this model, and this universe is found to be an isotropic one. Since the volume of the universe is not zero at $t=0$ it seems that dark energy existed in this model before the evolution of this era. In this model, though the modified scale co-variant gravitation theory seems to be a cause of the accelerated expansion of the universe, the dark energy here also enhances this expansion [see Fig. \ref{fig1}, Fig. \ref{fig2}]. Fig. [\ref{fig2a}] indicates that the energy flow from dark energy to dark matter.\\
The universe in case II is seen to be a model with accelerating expansion. But the rate of expansion is decreasing with time [see Fig. \ref{fig3}, Fig. \ref{fig4}]. In this case the interaction $Q$ [see Fig. \ref{fig4a}] is found to exist throughout the evolution of the universe which means that the dark energy prevails in this model all through its life. Here the state-finder parameters ${r,s}$ are found as $\lbrace{r=1,s=0}\rbrace$ if $\alpha=\frac{1}{2}$, which indicates that our universe happens to be a $\Lambda$CDM universe which is the most accepted form of a universe with dark energy.\\
Regarding the universe in case III it seems that dark energy has much significance with a huge density and pressure at the beginning of the evolution of this model. Also the density and pressure of the other matter contents of this universe tend to $\infty$ as $t\rightarrow 0$ [see Fig. \ref{fig5}, Fig. \ref{fig6}] which indicates that the model started with a big bang. Here interaction between the dark energy and other matters largely depend on the nature of the dark energy present in this universe. It is exceptionally high at  the beginning of the universe if the dark energy happens to be of quintessence type. It is also seen that the interaction is instigated, enticed and enhanced by the gauge function $\phi(t)$. At cosmic time $t=(\frac{1}{b_1})^{\frac{1}{b}}$ there will be no interaction between dark energy and the matter part of the universe, they are separately conserved. Thus at and around this time something unusual may happen in our universe. At this particular event of time the impact of the effect of dark energy on this universe will be very small; and the relation between the density and pressure of the fluid contained in this universe becomes $\rho=2p$. Therefore there may be a squeezing or contracting tendency of this universe at and around this time, tending towards a gravitational collapse. Also we see that at this time the dark energy either vanishes or remains dormant. Thus it seems that there may be some relation between dark energy and gravitational collapse. And the pressure and the density of the universe at this point of time seem to be some multiples or the functions of $\phi$. Thus it seems that the universe at this stage is controlled by $\phi$. Again in this model, as the squared speed of sound is positive, it (this universe) is supposed to be stable against small perturbations, and thus it has a stable evolution background, and therefore here we are studying under real astrophysical situations, and our model in this case may be taken as a realistic one. Since this model happens to be a universe which is most suited to and in agreement with modern observations it may be said that the present accelerated expansion depends wholly on the dark energy contained in this universe, as the modified gravitation theory in the form of scale co-variant theory has nothing to do with it (the expansion) here.\\
The model universe obtained in case IV is seen to begin with a big bang at $t=0$. Here since $\omega=\frac{-1}{3}$ the dark energy contained in this universe will be of the quintessence type which is one of the most accepted forms of dark energy. In this model the interaction between dark energy and other matters decreases with time. And the deceleration parameter is found to be positive and therefore our universe is decelerating initially but will accelerate at late time because of ``cosmic re-collapse" (Nojiri and Odintsov, 2003) \cite{Nojiri/2003}. At cosmic time $t=c_1^\frac{-1}{c_2}$ there is a singularity which may be taken as a bounce. Again if $c_1\rightarrow \infty$, it is seen that in this model $\rho_m$ and $p_m$ exist whereas $\rho_d$ and $p_d$ tend to zero [see Fig. \ref{fig7}, Fig. \ref{fig8}]; and this means that there may be an era in this model when there is no dark energy in it. Thus it may be concluded that in this era, either there is no accelerated expansion of the universe, or if it be there then the modified theory of gravitation will be responsible for the accelerated expansion. It is seen that the volume of this model universe increases with time though the rate of expansion decreases slowly. The density and pressure of the fluid contained in this model universe are seen to decrease with time which are natural for a realistic universe. However this model may have a big rip singularity at $t=\infty$. 

\textbf{Acknowledgments:} S.M. acknowledges Department of Science \& Technology (DST), Govt. of India, New Delhi,
for awarding Junior Research Fellowship (File No.
DST/INSPIRE Fellowship/2018/IF180676).

\end{document}